\documentclass[%
 aip,
 sd,%
 amsmath,amssymb,
 reprint,%
]{revtex4-1}

\usepackage{graphicx}% Include figure files
\usepackage{dcolumn}% Align table columns on decimal point
\usepackage{bm}% bold math
\begin{document}

%\preprint{}

\title{A new focal-plane 3D imaging method based on temporal ghost imaging} %Title of paper



\author{Zunwang Bo, Wenlin Gong*, Shensheng Han}
\email[]{zwbo@siom.ac.cn}
%\thanks{}
\affiliation{Key Laboratory for Quantum Optics and Center for Cold Atom Physics of CAS, Shanghai Institute of Optics and Fine Mechanics, Chinese Academy of Sciences, Shanghai 201800, China}


\date{\today}

\begin{abstract}
A new focal-plane three-dimensional (3D) imaging method based on temporal ghost imaging is proposed and demonstrated. By exploiting the advantages of temporal ghost imaging, this method enables slow integrating cameras have an ability of 3D surface imaging in the framework of sequential flood-illumination and focal-plane detection. The depth information of 3D objects is easily lost when imaging with traditional cameras, but it can be reconstructed with high-resolution by temporal correlation between received signals and reference signals. Combining with a two-dimensional (2D) projection image obtained by one single shot, a 3D image of the object can be achieved. The feasibility and performance of this focal-plane 3D imaging method have been verified through theoretical analysis and numerical experiments in this paper.
\end{abstract}


\maketitle %\maketitle must follow title, authors, abstract and \pacs

\section{Introduction}
In order to accuately perceive the physical world around us, various of 3D surface imaging technologies have been proposed, developed and commercialized in the past several decades \cite{geng2011,mcmanamon2012,nguyen2015}. Each 3D imaging technique has its own set of advantages and disadvantages when considering the key performance indexes such as accuracy, resolution, speed, cost, and reliability. From the perspective of system stability, hardware maturity and cost, some focal-plane 3D imaging technologies using standard framing cameras have attracted plenty of research attentions, such as structured-light 3D surface imaging\cite{geng2011}, monocular vision\cite{michels2005}, binocular vision\cite{cochran1992} and intensity-encoded 3D flash imaging\cite{tamburino1992,taboada1992}, etc. The principle of structured-light 3D surface imaging techniques is to extract the 3D surface shape based on the information from the distortion of the projected structured-light pattern. For close-range imaging, structured-light 3D imaging can achieve high accuracy, but its performance will degrade greatly in long-range imaging scenario. Both monocular vision and binocular vision take advantage of simple system, but its imaging accuracy is not enough for some application requiring depth information with high-resolution. Intensity-encoded 3D flash imaging converts time-of-fight of the pulse light to intensity information by a polarization and a Pockels cell, high-resolution 3D imaging with high speed can be achieved, but it is also less likely to be used for long-range applications, particularly those that require low size, weight, and power\cite{ayer1992}.

Ghost imaging is an interesting non-local imaging technique. It originates from researches on entangled photons and then proved to be able to realize by classical thermal light\cite{pittman1995,strekalov1995,bennink2002,zhang2005}. An important feature of ghost imaging is that it extracts information with high-resolution by correlation measurements between received signals and reference signals, yet the received signals have no resolution and the reference signals do not have any interaction with the object.\cite{cheng2004,cao2005,ferri2005,shapiro2012}. It manifests in the time domain as that a slow integrating 'bucket' detector without time-resolved capability is used to detect the temporal object, and its temporal structure can be reconstructed on the reference signal\cite{ryczkowski2016,devaux2016,chen2013} with high-resolution. This slow integrating detection model is suitable for framing cameras, which are commercially available among all the large array sensors. Based on this idea, we propose a new 3D imaging method via temporal ghost imaging (TGI) in the framework of sequential flood-illumination and focal-plane detection. This imaging process is mainly realized in two steps, 2D imaging by single-shot and focal-plane depth imaging by multiple-shot. The principle of focal-plane depth imaging is converting the measurements of pulse time-of-flight into the integrating time of each pixel, which is achieved by TGI between reference signals and received signals. The imaging scheme and reconstruction principle are presented. The feasibility and performance are also verified by numerical experiments in the paper.
\section{Imaging scheme}

\begin{figure}
\includegraphics[width=8.0cm]{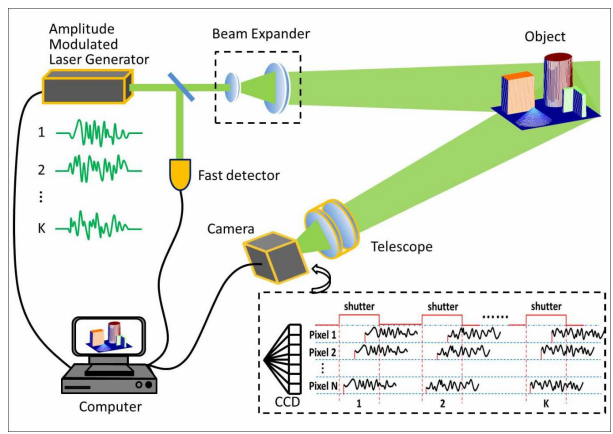}% Here is how to import EPS art
\caption{\label{fig:setup} Experimental schematic of focal-plane 3D imaging based on temporal ghost imaging.}
\end{figure}

Fig.~\ref{fig:setup} presents the fundamental scheme to implement focal-plane 3D imaging based on TGI. An amplitude modulated laser generator, which transmits laser pulses with random fluctuation in the time domain, is utilized as the illuminating source. The laser is first divided into two beams by a beam-splitter, namely reference beam and object beam. The reference beam is directly detected by a fast detector, which can record the intensity fluctuation of each laser pulse $I_{ref}^{i}(t)$, $t\in\left(P\right]$, $i=1,2\cdots K$, $K$ and $P$ is the number and width of the laser pulses respectively. Schematic diagram of the $K$ laser pulses has been shown in the upper left part of Fig.~\ref{fig:setup}. The object beam is shaped and expanded to make sure that the intensity distribution of the beam on the cross section is uniform. Laser pulses $I_{ref}^{i}(t)$, $i=1,2\cdots K$, reach the object in sequence and then be reflected back. A telescope is used to image the object onto the sensitive plane of the camera, which locates on the focal-plane of the receiving telescope. Since the surface of 3D object is non-planar along the optical propagation axis, reflected light of different points on the object surface will experience different flight time. Light reflected from the front parts of the object will reach the sensitive plane earlier than the back parts. Setting the delay time of the camera shutter relative to the laser emission and the shutter width to make that only part of the reflected light pulses can reach the camera before it shuts down. The insert drawing, which locates at the right bottom of Fig.~\ref{fig:setup}, shows the timing structure of the camera shutter and reflected pulses on different pixels. In this detection mode, the energy integrating time of each pixel is determined by the relative time-delay of the reflected pulses, but not the shutter width. It means that the depth information of the object is proportional to the energy integrating time of each pixel on the camera. The camera sequentially detects and records all the received light fields $I_{rec}^{i}(x,y)$, $i=1,2\cdots K$, $(x,y)$ presents the coordinate of each pixel on the camera. Amplitude modulated laser generator, fast detector and camera work synchronously during the whole measuring process. Based on the properties of orthogonality and ergodicity of the random intensity, the energy integrating time of each pixel can be separately calculated by the correlation coefficient of the reference signals with different integrating time and the received signals. For pixel $(x,y)$, the correlation coefficient, which is the function of the integration time, is defined by:
\begin{eqnarray}
&C(x,y,t')=\frac{\frac{1}{K}\sum\limits_{i=1}^K \Delta I_{rec}^{i}(x,y)\Delta I_{int}^{i}(t')}{\sqrt{\frac{1}{K^2}\left\{\sum\limits_{i=1}^K \left[\Delta I_{rec}^{i}(x,y)\right]^2\right\}\left\{\sum\limits_{i=1}^K \left[\Delta I_{int}^{i}(t')\right]^2\right\}}} \label{equ:a}
\end{eqnarray}
where $\Delta I^i=I^i-\frac{1}{K}\sum\limits_{i=1}^K I^i $ and
\begin{eqnarray}
I_{int}^{i}(t')=\int_{0}^{t'}I_{ref}^{i}(t)dt, t'\in \left(0,P\right) \label{equ:b}
\end{eqnarray}
is the integrating intensity of the reference signal with different integrating time. It can be predicted as that the correlation coefficient will obtain the maximum value when the integrating time of $I_{int}^{i}(t')$ is the same as the energy integrating time of the received signals $I_{rec}^{i}(x,y)$, so the energy integrating time of pixel $(x,y)$ can be achieved by:
\begin{eqnarray}
T(x,y)=\arg\max\limits_{t':t'>0} C(x,y,t') \label{equ:c}
\end{eqnarray}
Operating the computation pixel by pixel, the energy integrating time of each pixel $T(x,y)$ will be reconstructed. As the relative difference of $T(x,y)$ is proportional to the relative depth information of each pixel, the depth image of the object can be achieved by $R(x,y)=c\cdot T(x,y)/2$, where $c$ is the light speed. Combining the relative depth information by TGI and the 2D image by a single shot, the 3D image of the imaging object can be easily obtained.
\section{Simulation Experiment}
\begin{figure}
\includegraphics[width=8.0cm]{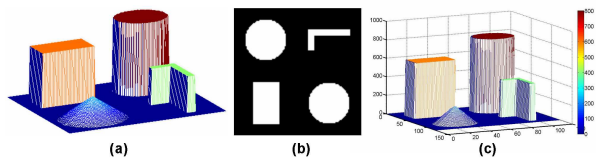}
\caption{\label{fig:result} Simulation result of focal-plane 3D imaging based on TGI.(a) is the original imaging object. (b) is the 2D image of the 3D object when looking from top to bottom, the gray information represents the object's reflectivity. (c) is the reconstructed depth image of the 3D object. The different colors of the images shown in (a) and (c) express different depth information.}
\end{figure}

The most important procedure of this new 3D imaging method is focal-plane depth imaging by TGI. Before implementing an experiment to validate its feasibility, a simulation is conducted firstly in this paper. In the numerical experiment, the amplitude modulated laser source is realized by 'rand' function in the MATLAB program, therefore its intensity fluctuation obeys uniform random distribution. The imaging object, as shown in Fig.~\ref{fig:result}(a), consists of four 3D geometric shapes, a cone, an 'L' shape object, a cuboid and a cylinder. The size of the imaging object in the length and width directions is $120\times120$ pixels, and the height of the four geometric shapes are $200$ pixels, $400$ pixels, $600$ pixels and $800$ pixels respectively. The length of the camera shutter is set as $1200$ pixels, which is long enough to accommodate all the returned pulses. Additive noise that may be caused by dark current or background radiation is taken into consideration. The measuring process can be described by equation $Y_{(x,y)}=AX_{(x,y)}+n_{(x,y)}$, where $X_{(x,y)}$ is a vector that presents the timing structure of the camera shutter and reflected pulses on pixel $(x,y)$, $A$ is the measurement matrix which concatenates the $K$ reference signals, $Y_{(x,y)}$ is the received signals on pixel $(x,y)$, $n_{(x,y)}$ is the additive white Gaussian noise. In the framework of ghost imaging, detection signal-to-noise (DSNR) is defined as the ratio of mean value of detection signals to noise variance. As the energy integrating time of each pixels is different, the DSNR will also be different from each other although the variance of noise on each pixel has little difference. In another word, pixels with more energy integrating time own higher DSNR. Fig.~\ref{fig:result}(b) is the 2D image of the object, which is used to determine the spatial distribution of the four geometric shapes with the same reflectivity information. The noise of the pixels that have no object can be easily filtered by that 2D image. Fig.~\ref{fig:result}(c) shows the depth information of the 3D object achieved by the reconstruction process with Eqs.~(\ref{equ:a})-~(\ref{equ:c}), in a condition of 6000 measurements and DSNR with $15dB$ corresponding to the maximum energy integrating time.

\begin{figure}
\includegraphics[width=8.0cm]{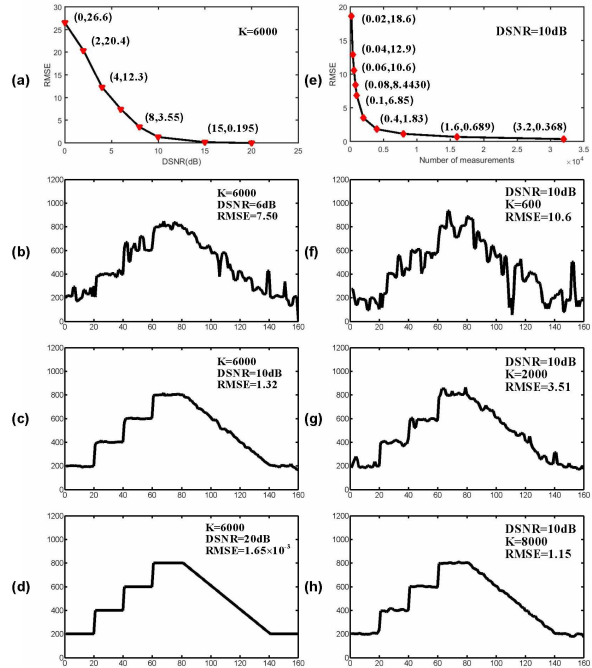}% Here is how to import EPS art
\caption{\label{fig:performance} Numerical experiments on performance of focal-plane depth imaging by TGI. (a)-(d) display the trend of reconstruction quality as the detection signal-to-noise (DSNR) varies from $0dB$ to $20dB$, with $6000$ measurements. (e)-(h) display the trend of reconstruction quality as the number of measurements varies from $200$ to $32000$, with a DSNR of $10dB$.}
\end{figure}

For testing the performance of the focal-plane 3D imaging method by TGI, numerical experiments with different measurements and different DSNR are carried out in Fig.~\ref{fig:performance}. In order to exhibit the reconstructed results more clearly, a one-dimensional (1D) object is utilized in the simulation. The maximum height of the object and the length of the camera shutter are also $800$ pixels and $1200$ pixels, respectively. Fig.~\ref{fig:performance}(a) presents the trend of reconstruction quality, which is described by root-mean-square error (RMSE) between the reconstructed depth information and the original object, as the DSNR varies from $0dB$ to $20dB$. The DSNR is defined relative to the maximum integrating energy, and the number of measurements in this numerical experiments is also $6000$. It is obvious that the imaging quality improves significantly with the increase of the DSNR. When the DSNR reaches a level of $20dB$, the proposed 3D imaging method nearly reconstructs the depth information of the object without any deviation. Fig.~\ref{fig:result}(b)-(d) separately displays the reconstructed results with the DSNR of $6dB$, $10dB$ and $20dB$. On the other hand, Fig.~\ref{fig:performance}(e) presents the curve of reconstructed quality as the number of measurements varies from $200$ to $32000$ with a DSNR of $10dB$. The reconstructed results imply that, if the number of measurements is lower than the Nyquist sampling theorem, the imaging quality improves dramatically as the measurements increase. Fig.~\ref{fig:result}(f)-(h) displays the corresponding results with the measurements of $600$, $2000$ and $8000$.

Another performance of this proposed 3D imaging method is its resolution, which includes the spacial-resolution of the 2D image and the range-resolution of the depth image. As known to all, the spacial-resolution of focal-plane imaging is determined by the numerical aperture of the receiving telescope and the pixel size of the camera. The depth-resolution achieved by TGI is determined by the upper value between the modulation bandwidth of the laser source and the response bandwidth of the fast detector. This problem has been experimentally demonstrated by the excellent work in paper [17].

\section{Discussion and Conclusion}
A focal-plane 3D imaging method based on temporal ghost imaging is proposed and demonstrated. This composite 3D imaging method is realized by a single shot to get the 2D image and multi-shots to get the depth image of the object, respectively. The most important procedure of depth image reconstruction is implemented by illumination with amplitude modulated laser source and detection with standard framing camera which locates on the focal-plane of a receiving telescope. This simple imaging framework not only increases the stability of the imaging system, but also reduces the cost. What's more, the integrating time of the reflected light can be changed by adjusting the width of laser pulse, the time-delay and shutter of the camera, therefore a long integrating time can be used to improve the influence of the camera's shot noise when it works in a low-flux condition. This focal-plane 3D imaging method may provide a new technology for application such as remote sensing, machine vision or facial recognition, etc.
\bibliography{aipsamp}% Produces the bibliography via BibTeX.

\end{document}